\input{aipcheck}

\documentclass[
    ,final            
  ]
  {aipproc}

\layoutstyle{6x9}
\usepackage{epsfig}

\begin{document}

\title{Fusion-Fission of $^{16}$O+$^{197}$Au at Sub-Barrier Energies}


\author{B.~B.~Back$^a$, C.~L.~Jiang$^a$, R.~V.~F.~Janssens$^a$, D.~J.~Henderson$^a$, B.~R.~Shumard$^a$, C.~J.~Lister$^a$, D.~Peterson$^a$, K.~E.~Rehm$^a$, I.~Tanihata$^{a,}$\footnote{Present address: TRIUMF, Vancouver, Canada}, X.~Tang$^a$, X.~Wang$^{a,b}$, S.~Zhu$^a$}{
  address={$^a$Argonne National Laboratory, Argonne, Illinois 60439, U.S.A.\\
$^b$University of Notre Dame, Notre Dame, Indiana 46556, U.S.A.}
}

\begin{abstract}
 The recent discovery of heavy-ion fusion hindrance at far sub-barrier energies has focused much attention on both experimental and theoretical studies of this phenomenon. Most of the experimental evidence comes from medium-heavy systems such as Ni+Ni to Zr+Zr, for which the compound system decays primarily by charged-particle evaporation. In order to study heavier systems, it is, however, necessary to measure also the fraction of the decay that goes into fission fragments. In the present work we have, therefore, measured the fission cross section of $^{16}$O+$^{197}$Au down to unprecedented far sub-barrier energies using a large position sensitive PPAC placed at backward angles. The preliminary cross sections will be discussed and compared to earlier studies at near-barrier energies. No conclusive evidence for sub-barrier hindrance was found, probably because the measurements were not extended to sufficiently low energies.

\end{abstract}

\maketitle


\section{Introduction}

Heavy-ion fusion processes, one of the main topics of this conference, are central to the understanding of nuclear reaction dynamics and extremely useful as a tool for populating high-spin states, the structure of which can be studied via their $\gamma$-decay cascade. Historically, the theoretical description of the fusion cross section has advanced in step with the availability of high quality experimental data for a wide range of fusion systems. This has led to a detailed understanding of the ion-ion potential determining the static interaction barrier and the consequences of the internal structure of the fusing nuclei as described in exquisite detail by coupled channels calculations. Recently, however, a detailed study of systems~\cite{Jiang}, for which the fusion excitation function has been measured down into the sub-microbarn region, have revealed a serious discrepancy between experimental data and the calculations, which does not appear to be reconcilable with the standard description. This effect has been a challenge for theory for several years. However, recently a new approach to calculating the ion-ion potential on the basis of a folding approach~\cite{Misicu}, which includes the compression energy, appears to point in the correct direction for an improvement of the theoretical description. The present work therefore represents a preliminary report on measurements of fission cross sections down to the micro-barn region for the $^{16}$O+$^{197}$Au system in order to provide quality data against which the improved theories can be tested.

\section{Experiment}
\begin{figure*}

\epsfig{file=Quad_PPAC_setup.eps,width=7cm}
\hspace{1cm}
\epsfig{file=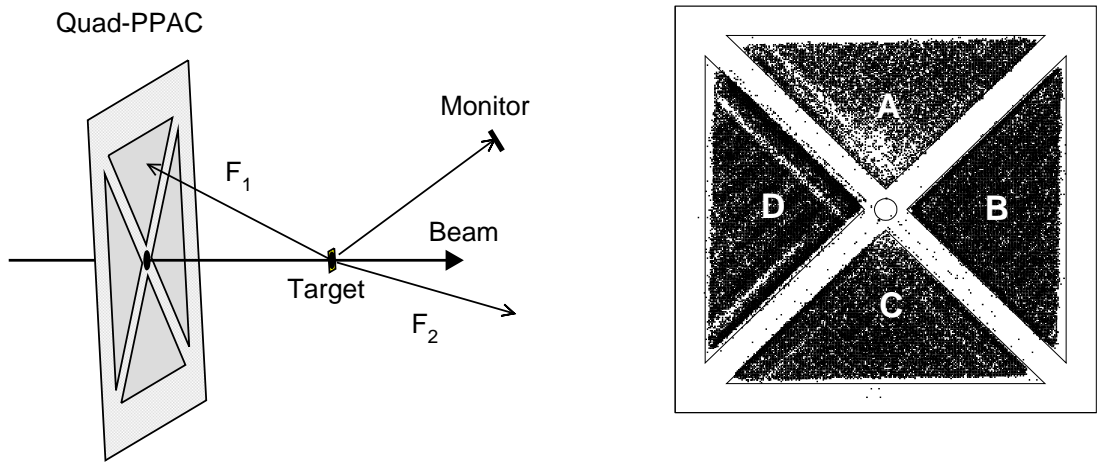,width=7cm}
\caption{Left panel: Schematic illustration of the experimental setup using a quad PPAC detector at backward angles. Right panel: Map of particle hits on the four segments (denoted {\it A, B, C,} and {\it D}) of the PPAC detector. Note that the PPAC detector is not 100\% efficient for $^{16}$O ions.}
\label{setup}
\end{figure*}

The present experiment was carried out at the ATLAS facility at Argonne National Laboratory. Fission cross sections were measured over energy $E_{\it lab}$ = 71.6 - 87.4 MeV (center-of-target) for the  $^{16}$O+$^{197}$Au system using the quad PPAC detector placed at backward angles as illustrated in the left panel of Fig.~\ref{setup}. This configuration was used in order to minimize the rate of elastically scattered $^{16}$O ions in the detector. Located at a distance of 18.3 cm from the target, each segment of the PPAC subtended a solid angle of 0.54 sr. The PPAC detector is position sensitive in two perpendicular directions as illustrated in a map of particle hits (elastics and fission fragments) shown in the right panel of Fig.~\ref{setup}. A Si monitor detector covering a solid angle of 0.098 msr was placed at 60$^\circ$ and used for cross section normalization.

\begin{figure*}[bth]
\epsfig{file=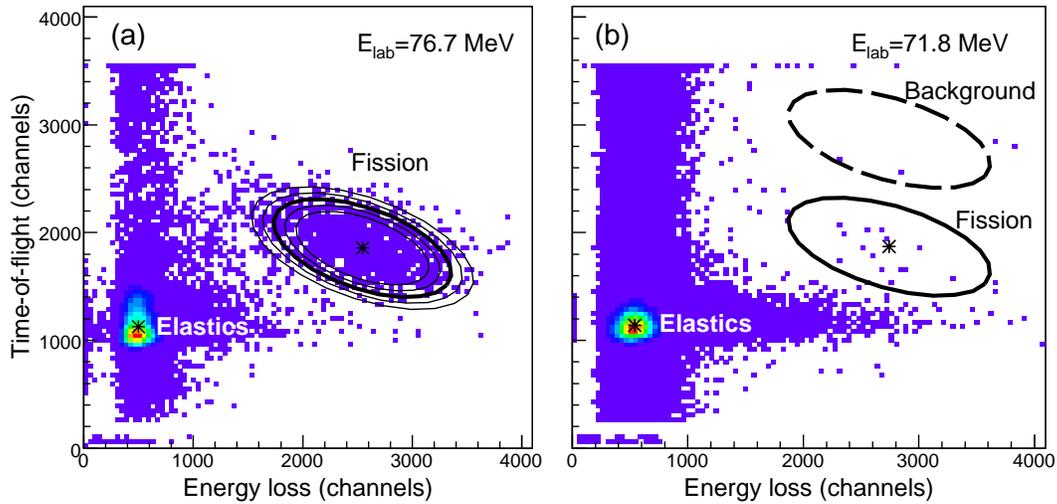,width=\textwidth}
\caption{Panel a: Energy loss versus time-of-flight contour map for $E_{\it lab}$=76.7 MeV for PPAC segment {\it B}. The centroids of the elastics and fission distributions are indicated by stars and five concentric ellipcal gates are used to integrate the number of fission fragments. Panel b: Same for $E_{\it lab}$=71.8 MeV. Solid and dashed ellipses indicate gate number three for fission and background, respectively.}
\label{etmap}
\end{figure*}

\begin{figure*}[hbt]
\epsfig{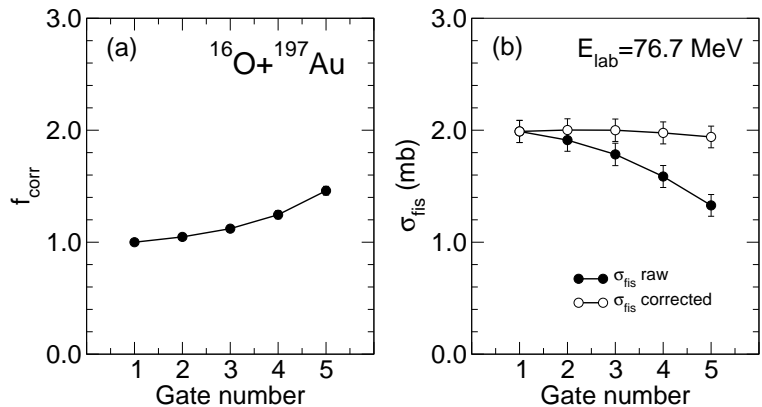}
\caption{Panel a: Correction factor shown as a function of gate number. Panel b: The uncorrected (solid circles) and corrected (open circles) cross sections are shown versus gate number for $E_{\it lab}$=76.7 MeV.}
\label{corfac}
\end{figure*}

The identification of fission fragments was based on the energy loss in the PPAC gas and the time-of-flight (from the target to the PPAC) utilizing the time structure of ATLAS beams. A map of the energy loss versus the flight-time is given in Fig.~\ref{etmap}a for E$_{\it lab}$= 76.7 MeV. Here the location of elastically scattered $^{16}$O ions and fission fragments is clearly visible. Five elliptical gates (solid curves) of linear dimensions of 100\%, 90\%, 80\%, 70\% and 60\% of the largest one are drawn around the fission fragment distribution. These gates are labeled 1 through 5 , respectively, with gate No. 1 being the largest one. 
At the lowest energies, the number of fission fragments is small and it may be difficult to draw the boundary for the fission fragment distribution. We have therefore developed a procedure which avoids the subjective judgment of the location of the fission fragments. From the data at energies above E$_{\it lab}$=75.5 MeV we determined the time difference (in channels)  between the centroid of the fission fragment distribution and that of the elastic peak as well as the ratio between the centroid of their energy loss. Also the average fraction of fission fragments falling within each of the five gates was determined from these runs. The corresponding correction factor $f_{\it corr}^i=N^i_{\it fis}/N^1_{\it fis}$, where $N^i_{\it fis}$ denotes the number of fission fragments falling within gate number $i$, is plotted as a function of gate number in Fig.~\ref{corfac}a.
 The solid points in Fig.~\ref{corfac}b illustrates the decrease in the calculated fission cross section obtained for PPAC segment {\it B} at $E_{\it lab}$=76.7 MeV as a function of gate number. The open circles, for which the average correction factor has been applied, shows that the calculated cross section is essentially independent of which gate is used. The present results use gate number three, which represents a compromise between including a large fraction of all fission fragments and excluding background events falling within the gate. 

At beam energies lower than E$_{\it lab}$=75.5 MeV we located the fission gate based on the measured centroid of the elastics peak using the time shift and energy loss ratio determined at higher energies. We also made a correction for background events, which were estimated on the basis of a gate that is shifted to longer flight-times as indicated by the dashed curve in Fig.~\ref{etmap}b. Even at the lowest energy, this correction amounts to only $\sim$25\%, and it decreases rapidly at higher energies. Finally, the preliminary results presented were obtained as an average over PPAC detector segments {\it A, B,} and {\it C}; segment {\it D} suffered some damage early in the experiment and was deemed unreliable for cross section determination. 

Since the present experiment measures only the cross section in the very backward angular range $\theta \sim 132^\circ - 174^\circ$, knowledge of the fission anisotropy is required in order to estimate the total fission cross section. We have used a polynomial fit to the measurements of Viola {\it et al}~\cite{Viola} to extrapolate it to the lower beam energies of the present experiment. Finally the absolute cross sections were obtained by normalization to Rutherford scattering observed in the monitor detector located at $\theta=60^\circ$. 

\section{Cross sections}
\begin{figure*}[bth]
\epsfig{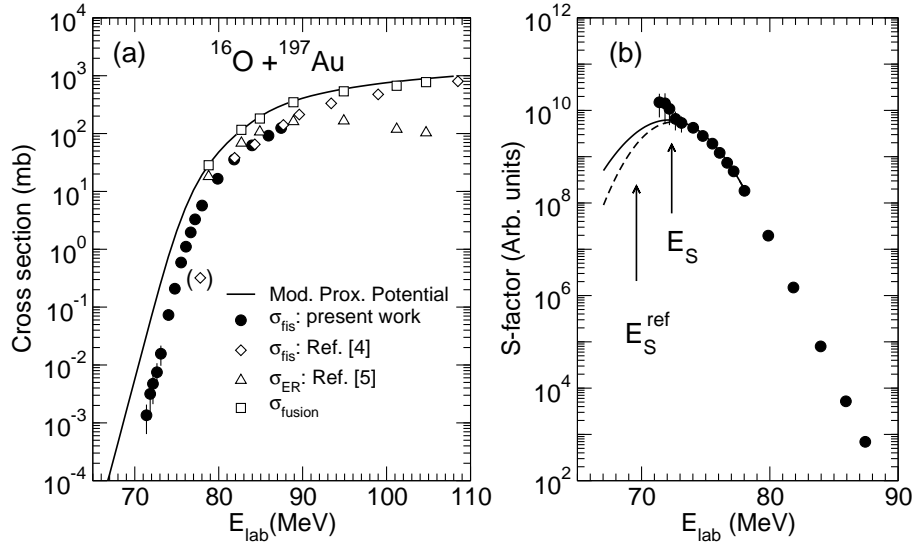}
\caption{Panel a: Present preliminary fission cross sections (solid circles) are compared to those of Sikkeland {\it et al.}~\cite{Sikkeland} (open diamonds). Also evaporation residue {\it et al.}~\cite{Brinkmann} (open triangles) and total fusion cross sections (open squares) are shown, the latter compared to predictions using a modified proximity potential~\cite{Back}. Panel b: S-factor representation of the present fission cross section. See text for further details.}
\label{crosec}
\end{figure*}

The preliminary results are shown as solid circles in Fig.~\ref{crosec}a and compared to earlier measurements by Sikkeland~\cite{Sikkeland} (open diamonds). The two measurements exhibit very good agreement over the overlapping energy range except for the lowest data point by Sikkeland (in parenthesis) which clearly deviates from the present data. 

Evaporation residues measured by Brinkmann {\it et al.}~\cite{Brinkmann} are shown as open triangles. When added to the present data we obtain the total fusion cross section (open squares) which agrees very well with a simple model estimate based on the modified proximity potential~\cite{Back}. 

Fig.~\ref{crosec}b shows the fission cross section in terms of the astrophysical $S$-factor, which is defined as $S=\sigma_{\it fis} E_{\it cm} \exp(2 \pi \eta)$, where $\eta$ is the Sommerfeld parameter. A maximum in the $S$-factor representation indicates the onset of sub-barrier hindrance~\cite{Jiang}. From the present data it is not obvious whether there is a maximum in the $S$-factor, but it is also not excluded as shown by the two curves that are obtained from fits to the logarithmic derivative of the cross section excluding the lowest data points. These fits, represented by solid and dashed curves in Fig.\ref{crosec}b, give an $S$-factor maximum at $E_s$. From the systematics~\cite{Jiang} of a large number of systems with sub-barrier hindrance, one expects that the hindrance should set in, and the maximum of the $S$-factor occur, at the energy indicated by the arrow labeled $E_{S}^{ref}$, $\it i.e.$ substantially lower in energy than probed in the present work. Judging from the present experience it may be very difficult to measure the fission cross section at this cross section level. However, as it appears that a large fraction of the total fusion cross section goes into the evaporation channel, a measurement of this exit channel may indeed reveal the sub-barrier hindrance in this system. 

\section{Summary}

A preliminary excitation function for the fission cross section in $^{16}$O+$^{197}$Au has been measured by detecting fission fragments at very backward angles. The results are in good agreement with those obtained by Sikkeland {\it et al.} (except for the lowest energy). Examination of the data, plotted as the astrophysical $S$-factor, does not reveal an obvious onset of hindrance. This behavior is not in conflict with the overall systematics, on the basis of which the hindrance onset is expected at an even lower enegy.


\begin{theacknowledgments}
This work was supported by the U. S. Department of Energy, Nuclear Physics Division, under contract No. W-31-109-ENG-38.
\end{theacknowledgments}

\end{document}